
\input harvmac.tex
\def\xhp{{\hat x^+}}
\def\xhm{{\hat x^-}}
\def\xhpm{{\hat x^\pm}}
\def\sq{{\vbox {\hrule height 0.6pt\hbox{\vrule width 0.6pt\hskip 3pt
   \vbox{\vskip 6pt}\hskip 3pt \vrule width 0.6pt}\hrule height 0.6pt}}}
\def\k{\kappa}
\def\a{\alpha}
\def\b{\beta}
\def\g{\gamma}
\def\d{\delta}
\def\r{\rho}
\def\f{\phi}
\def\tp{T_{++}^{\r,\f}}
\def\tm{T_{++}^{\rm M}}
\def\ppls{\partial_+}
\def\pmin{\partial_-}

\def\exp{{\rm exp}}

\lref\CGHS{C. Callan, S. Giddings, J. Harvey and A. Strominger,
Phys.~Rev.~{\bf D45}(1992) R1005.}
\lref\RST{J.~Russo, L.~Susskind and L.~Thorlacius, "Black Hole Evaporation
in $1+1$ Dimensions", Stanford preprint SU-ITP-92-4
(January 1992).}
\lref\BDDO{T.~Banks, A.~Dabholkar, M.~Douglas and M.~O'Loughlin, "Are
Horned Particles the Climax of Hawking Evaporation?",
Rutgers preprint RU-91-54 (January 1992).}
\lref\ST{L.~Susskind and L.~Thorlacius, "Hawking Radiation and Back-Reaction",
Stanford preprint SU-ITP-92-12 (March 1992).}
\lref\Hawk{S.~Hawking, "Evaporation of Two-Dimensional Black Holes",
Cal Tech preprint CALT-68-1774 (February 1992).}
\lref\BGHS{B.~Birnir, S.~Giddings, J.~Harvey and A.~Strominger, "Quantum
Black Holes", Chicago preprint EFI-92-16 (March 1992).}
\lref\Ghost{A.~Strominger, "Fadeev-Popov Ghosts and 1+1 Dimensional Black
Hole Evaporation", UCSBTH-92-18 (May 1992).}
\lref\DK{J.~Distler and H.~Kawai, Nuc.~Phys.~{\bf B321} (1989) 509.}
\lref\David{ F.~David, Mod.~Phys.~Lett. {\bf A3} (1988) 1651.}
\lref\TSEY{J.G. Russo and A.A. Tseytlin, "Scalar-Tensor Quantum Gravity in
Two Dimensions", SU-ITP-92-2/DAMTP-1-1992 (January 1992).}
\lref\CHAM{A. Chamseddine, Phys. Lett. {\bf B256} (1991)
379, Nucl. Phys. {\bf B368} (1992) 98;
T. Burwick and A. Chamseddine, ``Classical and Quantum
Considerations of Two-Dimensional Gravity", Z\"urich
preprint ZU-TH-4/92 (April 1992).      }
\lref\RSTE{ J.G. Russo, L. Susskind and L. Thorlacius, ``The Endpoint of
Hawking Radiation", Stanford preprint SU-ITP-92-17 (June 1992).}

\Title{\vbox{\baselineskip12pt
\hbox{PUPT-1320}\hbox{hepth@xxx/9205089}\hbox{includes Note Added 8/92}}}
{Liouville Models of Black Hole Evaporation}
\centerline{Adel Bilal{$^\dagger$}\footnote{*}
{\baselineskip8pt
On leave of absence from Laboratoire de Physique Th\'eorique de
l'Ecole Normale Sup\'erieure, 24 rue Lhomond, 75231 Paris Cedex 05,
France; Laboratoire Propre du CNRS.}~~ and
  ~~Curtis Callan\footnote{$^\dagger$}
{bilal@puhep1.princeton.edu,~~callan@puhep1.princeton.edu}}
\centerline{\it Department of Physics, Princeton University}
\centerline{\it Princeton, NJ 08544}
\vskip .3in
\centerline{\bf Abstract}
\smallskip

\vbox{
\baselineskip12pt

A renormalizable two-dimensional quantum field theory, containing
a metric, a dilaton and $N$ massless scalar matter fields, has
been proposed as a model for black hole evaporation. Essential ingredients
are a dilaton-dependent cosmological constant and a Polyakov action reflecting
the conformal anomaly. Previous work on this model has been
done in the large-$N$ (weak coupling) approximation and clear evidence for
Hawking radiation and its back-reaction on the metric has been seen.
There are, however, quantum consistency questions since
the original model was only designed to be a $c=26$ conformal field theory
in the weak coupling limit. In this paper we construct new theories,
differing from the old
only in the dilaton dependence of the cosmological constant,
and reducing to it in the weak coupling limit. They
 are exact $c=26$ conformal field theories and presumably
consistent frameworks for discussing this problem. We also study
the new theories with a change
in the Polyakov action proposed by Strominger with a view to eliminating
unphysical ghost Hawking radiation. The classical equations
of motion of the new theories are explicitly soluble, thus permitting an exact
analysis of both static solutions and dynamic scenarios. While the static
solutions are, by and large, physically reasonable, the dynamical
solutions include puzzling examples where wrong-sign Hawking radiation
is stimulated by allowing matter to fall into a static solution.
We indicate how the latter problem may be resolved in the full quantum theory.}
\Date{5/92}
\eject

\newsec{Introduction}

In a recent paper \CGHS\ a renormalizable two-dimensional quantum field
theory was proposed as an instructive model for the study
of black hole quantum mechanics. The starting point is the following
``string-inspired" classical action for a metric, a dilaton and a collection
of $N$ conformally coupled massless matter fields:
\eqn\one
{S= { 1 \over 2\pi}\int d^2 x\sqrt{-g}\left[e^{-2\phi}(R+4(\nabla\phi)^2
+4\lambda^2)
-\half\sum_1^N(\nabla f_i)^2\right]~.}
As long as the cosmological constant $\lambda$ is non-zero, this action
has a number of nontrivial solutions: First, there is
a vacuum solution in which the dilaton is linear in the
spatial coordinate, so that the quantum coupling strength, $e^{\phi}$,
goes to zero on one side of the world and to infinity on the other (we
refer to this as the linear dilaton vacuum, or LDV). Second, there are
static black hole solutions which approach the weak coupling limit of the
LDV at infinity. Finally, there are also explicit solutions describing the
formation of a black hole by the infall of an arbitrary pulse of
massless scalar matter.

The physics of this model is most easily analysed in
conformal gauge, defined by
\eqn\two
{g_{+-} =-\half e^{2\rho}~~g_{\pm\pm} = 0~~x^\pm=\tau\pm\sigma~~.}
In this gauge, the classical action reduces to
\eqn\three{S_N = {1 \over \pi}\int\ d^2\sigma\ \Bigl[e^{-2\phi}
(2\partial_+ \partial_-\rho - 4\partial_+\phi\partial_-\phi
+\lambda^2 e^{2\rho})
              + \half\sum\limits^N_{i=1} \partial_+ f_i\partial_-
f_i \Bigr]~}
and the equations of motion for the fields $\rho$, $\phi$ and $f_i$ must be
supplemented by two constraints (the equations of motion for the missing
components of the metric $g_{\pm\pm}$):
\eqn\four
{T_{\pm\pm}=  e^{-2\phi}\left(  4\partial_\pm\rho
\partial_\pm\phi-2\partial^2_\pm \phi\right) +
	\half\sum_1^N \partial_\pm f_i\partial_\pm f_i=0}
Note that the $f_i$ satisfy free field equations and influence the
metric-dilaton system only through the constraints.

To study Hawking radiation one of course has to quantize the above system.
The principal observation of \CGHS\ was that, since the matter fields are
free, their only quantum effect is through the conformal anomaly. This is
accounted for by adding the Polyakov term (multiplied by $N$ to reflect the
multiplicity of scalar fields) to the action and correspondingly
modifying the constraints. The results are particularly simple in conformal
gauge:
\eqn\five
{\eqalign{S_N = {1 \over \pi}\int\ d^2\sigma\ & \Bigl[e^{-2\phi}
(2\partial_+ \partial_-\rho - 4\partial_+\phi\partial_-\phi
+\lambda^2 e^{2\rho})\cr
             & + \half\sum\limits^N_{i=1} \partial_+ f_i\partial_-
f_i - {N \over 12} \partial_+\rho\partial_-\rho\Bigr]~,\cr
T_{\pm\pm} = e^{-2\phi}&(4\partial_\pm\phi\partial_\pm\rho -
2\partial^2_\pm\phi) + \half \sum\limits^N_{i=1}
	\partial_\pm f_i\partial_\pm f_i \cr
                  & - {N \over 12}\left(\partial_\pm\rho\partial_\pm\rho -
\partial^2_\pm\rho \right)~. }}
Strictly speaking, it is also necessary to give a quantum treatment of the
graviton-dilaton sector as well, and it is not obvious how to do that.
However, if $N$ is large, it seems reasonable to assume that matter
quantum effects dominate those of gravity and to proceed with a classical
treatment of the system summarized in \five. This line of investigation,
initiated in \CGHS, has been pursued by many
authors \refs{\RST,\BDDO,\Hawk,\ST}
and we shall assume that the reader is familiar with at least the
essentials of these papers. The basic result is that
the large-$N$ system correctly accounts for Hawking radiation and its
backreaction on the metric so long as $e^{2\phi}$\ (a measure of the
strength of purely gravitational quantum corrections) is smaller than a
certain critical value (itself of order $1/N$) where a singularity, signalling
the breakdown of the approximation, if nothing else,
must occur. Unfortunately, in the large-$N$ treatment of the formation and
subsequent evaporation of a black hole, the appearance of such singularities
seems to be inevitable.

Evidently, a full quantum treatment of the graviton-dilaton sector
of the theory will be needed in order to make further progress.
In this paper we will show how to obtain some exact results for theories
of the type given in \five . The most important issue concerns the
central charge: for quantum consistency, the theory must be a conformal
field theory with $c=26$ and there is no reason to believe that the
action \five\ does not require corrections to achieve this. By adapting
familiar 2d quantum gravity techniques \refs{\DK,\David},
we will be able to identify modifications of \five\ (having no effect
on the weak coupling behavior of the physical degrees of freedom)
which turn it into a $c=26$ conformal field theory. Since the resulting
theories turn out to be exactly soluble, we will be able to address the
question whether they have a satisfactory physical interpretation in terms
of the formation and evaporation of black holes.

\newsec{Transforming to Free Fields and Fixing the Central Charge}

With this background in mind, let us now turn to the problem of quantizing
the theory described by \five. This action has two pieces which we will
treat separately: all the terms involving derivatives of fields, which we
will collectively call $S_{\rm kin}$ and the cosmological constant term,
proportional to $\lambda^2$, which we will call $S_{\rm cos}$. Both require
some generalization: There is no reason to believe that the coefficient $N/12$
of the Polyakov term in $S_{\rm kin}$ is exact, so we will replace it by
$\kappa$, where $\kappa$ will be fixed by the $c=26$ requirement ($\kappa$
of course should reduce to $N/12$ in the large-$N$ limit). Alternatively,
one might argue \`a la David, Distler and Kawai \refs{\DK,\David}\
about the measure in the path integral and arrive at the same  conclusion.
Also, there is no reason to believe that the specific dilaton
dependence of $S_{\rm cos}$ specified in \five\ is exact. From string theory
experience we might expect it to be a more general power series in the
loop coupling constant squared $e^{2\phi}$ (the dependence specified in
\five\ is appropriate for a tree-level cosmological constant and should
be accurate in the weak coupling, or $e^{2\phi}\to 0$, limit only).

The generalized action of interest to us can therefore be written
as $S_\kappa=S_{\rm kin}+S_{\rm cos}$ where
\eqn\six{
\eqalign{S_{\rm kin} = {1 \over \pi}\int\ d^2\sigma\ \Bigl[e^{-2\phi}&
(2\partial_+\rho\partial_-\phi +2\partial_+\phi\partial_-\rho
	- 4\partial_+\phi\partial_-\phi)\cr
             & + \half\sum\limits^N_{i=1} \partial_+ f_i\partial_-
f_i - {\kappa} \partial_+\rho\partial_-\rho\Bigr]~,\cr
S_{\rm cos} = {1\over\pi}\int d^2\sigma &
		e^{-2\phi} \lambda^2 D(\phi)e^{2\rho}~~,}}
and all we know about $D(2\phi)$ is that it should approach unity in the weak
coupling limit.
Since the cosmological constant term does not affect the constraints,
they are the same as in \five\ with the replacement of $N/12$ by $\kappa$:
\eqn\sixone{\eqalign{
T_{\pm\pm} = e^{-2\phi}&(4\partial_\pm\phi\partial_\pm\rho -
2\partial^2_\pm\phi) + \half \sum\limits^N_{i=1}
        \partial_\pm f_i\partial_\pm f_i \cr
                  & - { \kappa}\left(\partial_\pm\rho\partial_\pm\rho -
\partial^2_\pm\rho \right)~. }}

At this point it is easy to see the origin of
the troubles that were found \refs{\RST,\BDDO,\Hawk} to afflict the
{large-$N$} limit (and perhaps the theory as a whole!):
The kinetic action density is of the form
$\partial_+\Phi\cdot M(\phi)\cdot\partial_-\Phi$ where $\Phi$ collects
the $N+2$ fields of the problem into a vector and $M(\phi)$ is an $(N+2)\times
(N+2)$ matrix. The determinant of $M$ has the value
\eqn\sixtwo{\det(M)= -4e^{-4\phi}(1-\kappa e^{2\phi}) \left( -\half\right)
^N}
and something singular obviously happens
at the critical value of the dilaton field, $\kappa e^{2\phi}=1$, where
the determinant changes sign. We will have more to say about how to deal
with this issue later on.

We will quantize this theory in two steps: First, we will show that $S_{\rm
kin}$
is actually a free field theory with an improvement term and that it has
$c=26$ if we choose $\kappa=(N-24)/12$. Then we will construct
$(1,1)$ operators relative to this simple conformal field theory and
so identify the correct cosmological constant function $D(\phi)$.
Our first main point is that $S_{\rm kin}$ can be reduced to a free theory
by a sequence of field redefinitions. Applying
\eqn\seven{\omega={e^{-\phi}\over\sqrt{\kappa}}\qquad
			\chi=\half(\rho+\omega^2)}
to the action \six\ and constraints \sixone\ gives
\eqn\eight{\eqalign{
S_{\rm kin}=&{1 \over \pi}\int\ d^2\sigma\ \Bigl[-4\kappa\partial_+\chi
	\partial_-\chi+4\kappa(\omega^2-1)\partial_+\omega\partial_-\omega
	+\half\sum\limits^N_{i=1} \partial_+ f_i\partial_-f_i \Bigr] \cr
S_{\rm cos}=&{1 \over \pi}\int\ d^2\sigma\ \Bigl[\kappa\lambda^2
	\omega^2 e^{-2\omega^2}D(1/\kappa\omega^2) e^{4\chi}\Bigr] \cr
T_{\pm\pm}=&-4\kappa\partial_\pm\chi\partial_\pm\chi+
				2\kappa\partial_\pm^2\chi
	+4\kappa(\omega^2-1)\partial_\pm\omega\partial_\pm\omega
		+ \half\sum\limits^N_{i=1} \partial_\pm f_i\partial_\pm f_i~.
}}
A further transformation on $\omega$ alone,
\eqn\nine{
\partial\Omega=\sqrt{\omega^2-1}~\partial\omega\ \ \Rightarrow\ \
	\Omega =\half\omega\sqrt{\omega^2-1}-
		\half \log(\omega + \sqrt{\omega^2 -1})~,}
finally reduces $S_{\rm kin}$ and $T_{\pm\pm}$ to free field form:
\eqn\ten{\eqalign{
S_{\rm kin}=&{1 \over \pi}\int\ d^2\sigma\ \Bigl[-4\kappa\partial_+\chi
        \partial_-\chi+4\kappa\partial_+\Omega\partial_-\Omega
        +\half\sum\limits^N_{i=1} \partial_+ f_i\partial_-f_i \Bigr] \cr
T_{\pm\pm}=&-4\kappa\partial_\pm\chi\partial_\pm\chi+
                                2\kappa\partial_\pm^2\chi
     +4\kappa\partial_\pm\Omega\partial_\pm\Omega
                 +\half\sum\limits^N_{i=1} \partial_\pm f_i\partial_\pm f_i~.
}}

There are various subtleties to discuss at this point. As is apparent from
the definition \seven\ of $\omega$, we are assuming that $\kappa>0$. We
shall stay with that assumption for the moment and come back later to the
question of what happens when $\kappa<0$. The principal issue concerns the
range over which the fields $\phi$, $\omega$ and $\Omega$ are supposed to
vary. As $\phi$ ranges from $-\infty$ to $\infty$, $\omega$ ranges from
$0$ to $\infty$. But \eight\ shows that the signature of the $\omega$
kinetic term changes sign at $\omega=1$. In the large-$N$ treatments of
this problem, a curvature singularity was always found at this critical
value of the dilaton field and it has been suggested that this is a natural
boundary and that $\omega$ should be restricted to range from $1$ to $\infty$
(the corresponding range of $\Omega$ is then from $0$ to $\infty$).
Such restrictions are certainly not natural from the point of view of the
action \ten\ and we have no idea how to implement them (or
even if they are really necessary). To proceed, we will assume that such
boundary conditions are either not necessary, or harmless, and that all
relevant information about the conformal properties of the theory is contained
in the action.

It is now a straightforward matter to analyze the central charge. The argument
is clearest if we rescale the fields $\chi$ and $\Omega$ in \ten\ by a factor
of $2\sqrt{\kappa}$ so as to eliminate $\kappa$,
obtaining
\eqn\eleven{\eqalign{
S_{\rm kin}=&{1 \over \pi}\int\ d^2\sigma\ \Bigl[-\partial_+\hat\chi
        \partial_-\hat\chi+\partial_+\hat\Omega\partial_-\hat\Omega
        +\half\sum\limits^N_{i=1} \partial_+ f_i\partial_-f_i \Bigr] \cr
T_{\pm\pm}=&-\partial_\pm\hat\chi\partial_\pm\hat\chi
                                +\sqrt{\kappa}\partial_\pm^2\hat\chi
    +\partial_\pm\hat\Omega\partial_\pm\hat\Omega
                 +\half\sum\limits^N_{i=1} \partial_\pm f_i\partial_\pm f_i~.
}}
The $N$ matter fields and the free $\hat\Omega$ field each of course
contribute unity to the central charge. The $\hat\chi$ field is free, but
has an improvement term, so its contribution is shifted from unity.
With our normalizations, one would normally set $c_\chi=1+12(\sqrt\kappa)^2>1$.
However, since $\hat\chi$ has a "wrong-sign" kinetic term (relative to
the matter fields), its central charge is {\it decreased} from unity by
the improvement term and the correct formula is $c_\chi=1-12(\sqrt\kappa)^2$.
The net result is
$	c= (1-12\kappa)~+1~+N = N~+2~-12\kappa$
and we can set $c=26$ by taking
\eqn\twelve{\kappa=(N-24)/12~.}
This behaves as expected
for large $N$, but it is perhaps a bit surprising that the $1/N$ corrections
stop at $O(N^0)$.

\def\chib{{\bar\chi}}
Finally, we have to ask whether the above line of argument can be carried
through for $\kappa<0$ so that we can deal with the case $N<24$. Since the
central charge formula is perfectly analytic in $N$, there should be no
difficulty, but it will be instructive to consider the question in detail.
The version of the field redefinitions of \seven\ appropriate to the
$\kappa<0$ case is
\eqn\thirteen{\bar\omega={e^{-\phi}\over\sqrt{|\kappa|}}\qquad
			\chib=\half(\rho-\bar\omega^2)~.}
Their application to the action \six\ and constraints \sixone\ gives
\eqn\fourteen{\eqalign{
S_{\rm kin}=&{1 \over \pi}\int\ d^2\sigma\ \Bigl[4|\kappa|\partial_+\chib
\partial_-\chib-4|\kappa|(\bar\omega^2+1)\partial_+\bar\omega\partial_
-\bar\omega
	+\half\sum\limits^N_{i=1} \partial_+ f_i\partial_-f_i \Bigr] \cr
S_{\rm
cos}=&{1 \over \pi}\int\ d^2\sigma\ \Bigl[|\kappa|\lambda^2
	\bar\omega^2 e^{2\bar\omega^2}D(1/|\kappa|\bar\omega^2)
e^{4\chi}\Bigr] \cr
T_{\pm\pm}=&4|\kappa|\partial_\pm\chib\partial_\pm\chib-
				2|\kappa|\partial_\pm^2\chib
	-4|\kappa|(\bar\omega^2+1)\partial_\pm\bar\omega\partial_\pm\bar\omega
		+ \half\sum\limits^N_{i=1} \partial_\pm f_i\partial_\pm f_i~.
}}
Two important things have changed compared to the previous case: the signature
of the $\chi$ kinetic term has switched from time-like to space-like
(so that its improvement term now {\it increases} the central charge) and
the signature of the $\bar\omega$ term is now definite (and timelike). A final
transformation on $\bar\omega$ alone,
\eqn\fifteen{
\partial\bar\Omega=-\sqrt{\bar\omega^2+1}~\partial\bar\omega \ \ \Rightarrow\
\ \
	\bar\Omega =-\half\bar\omega\sqrt{\bar\omega^2+1}-
		\half \log(\bar\omega + \sqrt{\bar\omega^2 +1})~,}
transforms the action and constraints to free field form:
\eqn\sixteen{\eqalign{
S_{\rm kin}=&{1 \over \pi}\int\ d^2\sigma\ \Bigl[\partial_+\hat{\bar\chi}
        \partial_-\hat{\bar\chi}-\partial_+
		\hat{\bar\Omega}\partial_-\hat{\bar\Omega}
        +\half\sum\limits^N_{i=1} \partial_+ f_i\partial_-f_i \Bigr] \cr
T_{\pm\pm}=&\partial_\pm\hat{\bar\chi}\partial_\pm\hat{\bar\chi}-
                                \sqrt{|\kappa|}\partial_\pm^2\hat{\bar\chi}
       - \partial_\pm\hat{\bar\Omega}\partial_\pm\hat{\bar\Omega}
                + \half\sum\limits^N_{i=1} \partial_\pm f_i\partial_\pm f_i~.
}}
(we have rescaled the $\chi$ and $\Omega$ fields by a further factor of
$2\sqrt{|\kappa|}$ in order to have standard normalization).

The central charge calculation is the same as before, except that, since the
$\chi$ field now has spacelike signature, the improvement term {\it increases}
its central charge. With this modification, we have
$c= (1+12|\kappa|)~+1~+N = N~+2~-12\kappa$
and the $c=26$ condition reduces to $|\kappa|=(24-N)/12$. Since $\kappa$ is
negative, this is the same result as before.
We note that the kinetic part of the action and
constraints in \sixteen, when written in terms of
the unrescaled fields $\chib$ and $\bar\Omega$, are exactly the same as
\ten\ for $\chi$ and $\Omega$, and the dependence on $\kappa$ is analytic.
One interesting difference
concerns the ranges of the fields. Referring back to the discussion around
\sixtwo , we see that if $\kappa<0$, there is no real value of the dilaton
field at which the determinant vanishes and therefore no singularity to avoid
by restricting the range of variation of any of the fields. It is true that
as $\phi$ ranges from $-\infty$ to $\infty$, $\omega$ and $\Omega$ only take
on positive values. Since $\omega$ is $1/g_{\rm string}$,
and since all physical
quantities depend only on $g_{\rm string}^2$, it seems quite reasonable
to let the $\omega$ fields range over negative values as well.

\newsec{Constructing $(1,1)$ Operators}

The next task is to identify the cosmological constant actions, $S_{\rm cos}$,
which can consistently be added to the $c=26$ conformal theory defined by
$S_{\rm kin}$. The {\it infinitesimal} conformal perturbations are identified
by the operators of conformal weight $(1,1)$ with respect to $S_{\rm kin}$.
Finding the exact finite conformal perturbations is a lot harder, but
experience with Liouville-like theories (and the theory at hand is of
that general type, as we shall see) shows that the infinitesimal perturbations
are usually exact (no higher-loop corrections to the anomalous dimension
operator). We don't know for sure that this is so, but, for the purposes of
this paper, we will assume that it is enough to identify the (1,1) operators.

Consider first the $\kappa>0$ case. It is of course easiest to construct the
conformal weight operator in terms of the conventionally normalized free
fields $\hat\chi$ and $\hat\Omega$ (using the action and energy-momentum
tensor specified in \eleven). The condition for a scalar function
$V(\hat\chi,\hat\Omega)$ to be a $(1,1)$ operator is
\eqn\seventeen{
\Bigl[\half\sqrt{\kappa}{\partial\over\partial\hat\chi}+
	{\partial^2\over\partial\hat\chi^2}-
	{\partial^2\over\partial\hat\Omega^2}\Bigr]V(\hat\chi,\hat\Omega)=
		V(\hat\chi,\hat\Omega)~~.}
The opposite signs of the second derivative terms of course stems from the
opposite signatures of the two fields. The linear term in
$\partial/\partial\hat\chi$ of course comes from the improvement term
and its normalization can be checked in the classical limit
($\kappa\to\infty$) where the dimension operator is dominated by the
$\sqrt{\kappa}$ term: The classical cosmological constant term displayed
in \eight\ can be rewritten as
\eqn\eighteen{
V_{\rm class}=\kappa\omega^2e^{-2\omega^2}e^{4\chi}=
	F(\hat\Omega)e^{{2\over\sqrt{\kappa}}\hat\chi}}
and it is obviously a solution of the ``classical" $(1,1)$ condition
\eqn\nineteen{
\half\sqrt{\kappa}{\partial\over\partial\hat\chi} V(\hat\chi,\hat\Omega)=
		V(\hat\chi,\hat\Omega)~~.}
There are many exact solutions to \seventeen\ and a particularly simple
class is given by (the normalization is chosen for later convenience)
\eqn\twenty{
V_{\rm exact}^\pm={\kappa\over 4 e} e^{\pm{2\over\sqrt{\kappa}}\hat\Omega}
		e^{{2\over\sqrt{\kappa}}\hat\chi}
={\kappa\over 4 e} e^{4\chi \pm 4 \Omega }~~.}
We are most interested in solutions which reduce to the classical cosmological
constant $e^{2(\rho-\phi)}$ in the classical limit $e^{\phi}\to 0$. The
solution
\twenty\ with the minus sign in the exponent in fact has this property. Using
the transformations given in \seven\ and \nine\ we find that
\eqn\twentyone{V_{\rm exact}={\kappa\over 4 e}e^{4\chi - 4 \Omega}=e^{2\rho}
e^{-2\phi} D(e^{2\phi})}
where
\eqn\twentytwo{
D(e^{2\phi})={1\over 4}(1+\sqrt{1-\kappa e^{2\phi}})^2
               \exp\Bigl[{(1-\sqrt{1-\kappa e^{2\phi}})
		\over(1+\sqrt{1-\kappa e^{2\phi}})}\Bigr] ~~.}
The suggestion is that with this choice of the cosmological constant function,
the theory defined in \six\ (with the appropriate value of $\kappa$) is
an exact $c=26$ conformal field theory. Though there are other
choices of $D(e^{2\phi})$ for which the theory is conformal, \twentytwo\ is the
unique one which reduces to the original classical action in the weak
coupling limit, i.e. $D(0)=1$.
Thus this exact conformal theory should have the same
structure of asymptotic states as were encountered in the original
semiclassical studies of these theories and should be suitable for the
exact study of the black hole evaporation problem. Note that the
cosmological constant function now has a branch point at precisely
the value of the dilaton field  where the determinant of the kinetic term
changes sign. We will have some comments on how to deal with the associated
singularities later on in the paper.

It remains to determine the function $D(\phi)$ appropriate to the $\kappa<0$
case. As the discussion of the central charge made clear, one can, for all
practical purposes, take the $\kappa>0$ formulas and analytically continue
them to $\kappa<0$. The $(1,1)$ operator with the desired weak coupling
behavior is easily seen to be
\eqn\twentythree{\eqalign{
\bar V_{\rm exact} = & {|\kappa|\over 4 e} e^{4\chib-4\bar\Omega}\cr
		=&{|\kappa|\over 4 e}
e^{4\chib}~(\bar\omega+\sqrt{1+\bar\omega^2})^2
	~e^{2\bar\omega\sqrt{\bar\omega^2+1}}\cr
		=&e^{2(\rho-\phi)}~
{1\over 4}(1+\sqrt{1+|\kappa|e^{2\phi}})^2
	\exp\Bigl[
	{(1-\sqrt{1+|\kappa|e^{2\phi}})\over(1+\sqrt{1+|\kappa|e^{2\phi}})}
	\Bigr]\cr
=& e^{2\rho -2 \phi} \bar D(e^{2\phi})	~.}}
An important point is that this function is well-behaved for all values of
$\phi$. Since, as we showed in the previous section, the kinetic action
is free of singularity as well, it seems plausible that the $\kappa<0$ theory
is a completely well-defined quantum theory of black hole evaporation.
We will explore this notion in a later section. Note also that,
in the strong coupling limit ($e^\phi\to \infty$), \twentythree\ asymptotes
to  ${|\kappa|\over 4 e} e^{2\rho}$ and the dependence of the action \six\
on the metric variable $\rho$ reduces to the standard Liouville action.
Since the Liouville theory is well-behaved and since the whole issue
of singularities has to do with the behavior of the theory at strong coupling,
this suggests that the $\kappa<0$ theory should have no singularity
problems at the quantum level.

\newsec{Exact Solutions}

We would now like to see what can be said about the
formation and subsequent evaporation of black holes in the conformally
invariant dilaton-gravity theory constructed in the preceeding sections.
Ideally, of course, we should develop a full quantum treatment.
While this may be possible, it has yet to be done and we will limit ourselves
here to a study of the classical solutions of the action \six\ (with the
special values of $\kappa$ and $D(\phi)$ described above). Apart from the
``improved" choice of $D(\phi)$, this is the approach adopted in all previous
work on this subject \refs{\CGHS,\RST,\BDDO} and, as has been explained
elsewhere, it amounts to a semiclassical treatment in which matter quantum
loops are accounted for, but graviton-dilaton loops are not. Such an approach
includes the basic physics of Hawking radiation, in which
left-moving infalling matter creates, via the anomaly, right-moving outgoing
matter radiation along with an appropriate back-reaction on the metric. The
approximation will of course fail if the dilaton-gravity theory becomes
strongly coupled anywhere in spacetime. We will defer discussion of the
self-consistency of the approximation until after we have constructed
the solutions of the classical equations. As we will now explain, with the
new conformally invariant form of $D(\phi)$ (\twentytwo\   or \twentythree),
the equations of motion become exactly soluble.

Consider the $\kappa>0$ case. In the previous sections we have shown that,
with the special choice \twentytwo\ for $D(\phi)$, the action \six\ has a
very simple form when expressed in terms of the fields $\chi$ and
$\Omega$ (these fields can be expressed in terms of the original
$\rho$ and $\phi$ fields using \seven\ and \nine :
\eqn\twentyfour{ S^{\r,\f} = {4\kappa\over\pi}\int d^2\sigma\bigl[
-\partial_+\chi
        \partial_-\chi+\partial_+\Omega\partial_-\Omega
+{\lambda^2\over 16 e} e^{4(\chi-\Omega)}\bigr]~.}
Things are even simpler in terms of $\Psi_\pm=\chi\pm\Omega$:
\eqn\twentyfive{ S^{\r,\f} = {4\kappa\over\pi}\int d^2\sigma\bigl[
-	\partial_+\Psi_+\partial_-\Psi_- +
	{\lambda^2\over 16 e} e^{4\Psi_-}\bigr]~.}
The equations of motion that follow from this action are simply
\eqn\twentysix{
 \partial_+\partial_-\Psi_-=0,\qquad \partial_+\partial_-\Psi_+=
	-{\lambda^2\over 4e} e^{4\Psi_-}~.}
For $\kappa < 0$ everything works out the same way except that now
\eqn\twentysixprime{
\partial_+\partial_-\Psi_-=  0, \qquad
\partial_+\partial_-\Psi_+=
+{\lambda^2\over 4e} e^{4\Psi-}~,}
where $\Psi_\pm=\bar\chi\pm\bar\Omega$.
The most general solution for $\kappa>0$ or $\kappa<0$
is easy to write down:
\eqn\twentyseven{\eqalign{
	2\Psi_-= & \alpha(x^+)+\beta(x^-)+
		\log(2/(\lambda\sqrt{|\kappa|})) +\half \cr
	2\Psi_+= & 2\gamma(x^+)-\alpha(x^+)+2\delta(x^-)-\beta(x^-)
		+\log(2/(\lambda\sqrt{|\kappa|})) +\half \cr
	\quad &  -{2\over\kappa}
		\int^{x^+}dy\ e^{2\alpha(y)}\int^{x^-}dz\ e^{2\beta(z)}~,}}
where $\alpha$ etc. are arbitrary functions of integration.
Since the $\Psi_\pm$ are known
functions of the original fields $\rho$ and $\phi$, the latter can
be found by solving a pair of transcendental equations:
\eqn\twentynine{\eqalign{
2\Omega(\f)=\log({1-\sqrt{1-\kappa e^{2\phi}}\over\kappa})-\phi+&
{e^{-2\phi}\over\kappa}\sqrt{1-\kappa e^{2\phi}} +\half\log|\kappa| =\cr
\gamma(x^+)+\delta(x^-)-&\alpha(x^+)-\beta(x^-)
	-{1\over\kappa}\int^{x^+}dy~e^{2\alpha(y)}\int^{x^-}dz~e^{2\beta(z)}\cr
\rho + \log(\lambda) +{e^{-2\phi}\over\kappa} =
	\gamma(x^+)+\delta(x^-)& - \half\log|\kappa| \cr
	& +\log(2)+\half-{1\over\kappa}\int^{x^+}dy~e^{2\alpha(y)}
	\int^{x^-}dz~e^{2\beta(z)}~.}}

One may choose any sufficiently regular functions $\alpha,\ldots,\delta$
in \twentyseven\ or \twentynine\ (though some restrictions
follow from boundary conditions of various kinds). Now, the equations of motion
are conformally invariant if $\f$ is a scalar and $\r$ transforms as
$\tilde\r(y)=\r(x)-\half\log({dy^+\over dy^-}{dy^+\over dy^-})$. Thus, two
sets of functions $\tilde\alpha(y),\ldots,\tilde\delta(y)$ and
$\alpha(x),\ldots,\delta(x)$ correspond to the same solution if they are
related by
\eqn\twentyeight{\eqalign{
\tilde\alpha(y^+)=\alpha(x^+)-\half\log{dy^+\over dx^+}~~&,~~
\tilde\beta(y^-)=\beta(x^-)-\half\log{dy^-\over dx^-} \cr
\tilde\gamma(y^+)=\gamma(x^+)-\half\log{dy^+\over dx^+}~~&,~~
\tilde\delta(y^-)=\delta(x^-)-\half\log{dy^-\over dx^-}~~.}}
Alternatively, these are just the rules for the transformation of
$\alpha,\ldots,\delta$ under a conformal coordinate transformation.

The $T_{\pm\pm}$ constraints associated with the dilaton-gravity action
have a particularly simple form in terms of the
$\chi$ and $\Omega$ fields (see \ten\ and \sixteen) and therefore
also in terms of the $\Psi_\pm$ fields. Evaluated on the solution \twentyseven,
the $\rho,\phi$-part of the constraints are
\eqn\thirty{\eqalign{
T_{++}^{\rho,\phi} =& \kappa(\partial_+(\gamma-\alpha))^2-
	\kappa(\partial_+\gamma)^2 + \kappa\partial_+^2\gamma\cr
T_{--}^{\rho,\phi} =& \kappa(\partial_-(\delta-\beta))^2-
	\kappa(\partial_-\delta)^2 + \kappa\partial_-^2\delta~. }}
They of course automatically satisfy the conservation equations
$\partial_\mp T_{\pm\pm}^{\rho,\phi}=0$.
As appropriate, we will try to interpret
non-zero values for $T_{\pm\pm}^{\rho,\phi}$
in terms of ingoing or outgoing fluxes
of the $f$-matter fields (i.e. $T^{\rm M}_{\pm\pm}\ne 0$).

A signficant advantage of having this exact solution, as compared to the
situation in the $D(\phi)=1$ theory, is that we can make
analytic statements about the global spacetime structure of solutions
without having to resort to numerical integration. One might wonder whether
the $D(\phi)=1$ equations could also be solved exactly. If so, there should
be other conserved quantities besides $T_{\pm\pm}$. We have checked that
no conserved quantities of dimensions one, three or four (except for trivial
ones built out of $T$ itself) exist. This strongly suggests that the
$D(\phi)=1$ equations are not integrable.

\newsec{Decoupling the Ghosts}

We now want to show that the procedures developed above can be applied to
an interesting variant of the action \six. This is important
because, as we shall see, the solutions of the theory we have been studying
are physically worrisome for a variety of reasons.
In this section, we will study a generalization of \six\ which can also be
treated using the techniques presented in this paper and whose
static solutions seem
to be physically perfectly satisfactory (at least for $N<24$).

The starting point of our procedure is a split of the action into a kinetic
part and a cosmological constant part. The kinetic part is the standard
dilaton-gravity kinetic term plus a Polyakov term representing the net
anomaly due to $N$ matter fields plus the dilaton-gravity fields. We adjust
constants so that, in conformal gauge, this kinetic action is a conformal
field theory and then construct the cosmological constant action as a $(1,1)$
operator with respect to the kinetic action. The resulting theory differs
from the original model only in the precise form of the cosmological constant
(and is qualitatively different only in the strong coupling region).

Strominger \Ghost\ has observed that there is good physical reason to consider
a modification of the kinetic action itself. His basic point is that
the form of the Polyakov anomaly action depends on the metric used to define
the path integral measure and that it is not obvious that one wants to use
the same metric for the graviton-dilaton-ghost system as one uses for the
matter fields. One could, as one often does in string theories, use a
Weyl-transformed metric of the type $g^{[\alpha]}_{ij}=e^{-\alpha\phi}g_{ij}$
(where $g_{ij}$ is the metric appearing in the graviton-dilaton action and
$\alpha$ is some constant). The action used so far in this paper corresponds
to using the ``true" ($\alpha=0$) metric to define all the measures.
This appears to make matter, ghosts, gravitons and dilatons all contribute
to the Hawking radiation on the same footing
(with the ghosts of course giving a negative flux!). This seems unphysical,
since the graviton and dilaton represent non-propagating degrees of freedom
and the ghosts should never appear in any on-shell process.

Strominger proposes a simple
way out of the problem: Use the rescaled metric $g^{[2]}_{ij}$ to define
the measure of the graviton-dilaton-ghost system. It has the nice
property that it is flat in {\it any} solution of the
($D(\f )=1$) classical theory \five\
(in conformal gauge $g_{ij}=e^{2\rho}\delta_{ij}$ and one can further gauge
fix a solution to $\rho=\phi$), thus decoupling the unphysical fields from
the geometry, at least in some leading order. In conformal gauge, this is
implemented by building the graviton-dilaton-ghost Polyakov term out of
$\rho-\phi$, while building the matter anomaly term out of $\rho$, as before.
The resulting kinetic action is a simple modification of \six:
\eqn\thirtyone{\eqalign{
S^\prime_{\rm kin} = {1 \over \pi}\int\ d^2\sigma\ \Bigl[e^{-2\phi}&
(2\partial_+\rho\partial_-\phi +2\partial_+\phi\partial_-\rho
	- 4\partial_+\phi\partial_-\phi)\cr
             & + \half\sum\limits^N_{i=1} \partial_+ f_i\partial_- f_i -
{N\over 12} \partial_+\rho\partial_-\rho
	+2 \partial_+(\rho-\phi)\partial_-(\rho-\phi) \Bigr]~.}}
We could have written the two types of Polyakov term with arbitrary
coefficients, to be determined by the requirement that the action
generate a $c=26$ conformal field theory, but we have anticipated the result
of that calculation. We will now show that this theory can be explicitly
transformed to free fields, that the $(1,1)$ cosmological constant
term can be written down explicitly, and that explicit exact solutions
of the classical equations of motion can be found. The argument is a simple
modification of what was presented in earlier sections, so we shall be brief.
(As a side remark, we note that all this could have been done for an
action with the anomaly term built out of the most general linear
combination of $\partial\r\partial\r$,  $\partial\f\partial\f$ and
$\partial\r\partial\f$. Since we have no physical motivation for this, we
will restrict our attention to \thirtyone.)

By analogy to \seven, we define new fields $\omega$ and $\chi$ by
\eqn\thirtytwo{
\omega = e^{-\phi}\qquad \rho = 2\chi + f(\omega)}
(note that this time we do not rescale $\omega$ by $\sqrt{\kappa}$)
and try to choose $f(\omega)$ so as to diagonalize the action. One can easily
show that $f(\omega)={2\over\kappa}(\log\omega-\omega^2/2)$
does the trick (where $\kappa=(N-24)/12$ as before).
The diagonalized kinetic action of the graviton-dilaton sector is
\eqn\thirtythree{
S^\prime_{\rm kin} = {1 \over \pi}\int d^2\sigma \Bigl[
{4\over\kappa}(\omega^2-(\kappa+2)+{\kappa+2\over 2\omega^2})
	(\partial_+\omega\partial_-\omega)
	-4\kappa(\partial\chi_+\partial\chi_-)\Bigr]~~.  }
If we define a new field $\Omega$ such that
\eqn\thirtyfour{
(\partial\Omega)^2={1\over\kappa^2}
	(\omega^2-(\kappa+2)+{\kappa+2\over 2\omega^2})(\partial\omega)^2\ \
	\Rightarrow\ \ \Omega = {1\over\kappa}
	\int^\omega dy\sqrt{y^2-(\kappa+2)+{\kappa+2\over 2y^2}} ~,}
then the kinetic action and constraints finally simplify to
(we do not write the matter contributions explicitly)
\eqn\thirtyfive{\eqalign{
S^\prime_{\rm kin}=&{1 \over \pi}\int\ d^2\sigma\ \Bigl[-4\kappa\partial_+\chi
        \partial_-\chi+4\kappa\partial_+\Omega\partial_-\Omega \Bigr] \cr
T_{\pm\pm}=&-4\kappa\partial_\pm\chi\partial_\pm\chi+
                                2\kappa\partial_\pm^2\chi
        +4\kappa\partial_\pm\Omega\partial_\pm\Omega~~. }}

This is precisely the same as the system \ten\ which we have already studied
and we can immediately draw the same conclusions: $S^\prime_{\rm kin}$
defines a $c=26$ conformal theory if $\kappa=(N-24)/12$ (which we have already
assumed) and the $(1,1)$ conformal fields are
$V_{(1,1)}^\pm = e^{4(\chi\pm\Omega)}$. It again turns out that the
appropriately normalized
$V_{(1,1)}^-$ reduces to the original classical cosmological constant
in the weak coupling ($\phi\to -\infty$) limit. This normalization depends
on the choice of integration constant in the integral for $\Omega$. To be
specific, we choose it such that in the weak coupling limit ($\omega\to\infty$)
$2\Omega$ is given by
\eqn\thirtyfiveone{
2\Omega\sim {e^{-2\phi}\over\kappa}+{\kappa +2\over 2}\f -\log 2
	+\half + \half\log|\kappa|+ {\kappa +2\over 8}e^{2\f}+O(e^{4\f})~.}
Then the correctly normalized conformally invariant cosmological constant reads
\eqn\thirtyfivetwo{
S_{\rm cos} = {\lambda^2\over\pi}\int d^2\sigma{|\kappa|\over 4e}
	e^{4\chi-4\Omega}~~}
and the full action to be solved,
the equations of motion and the general solution are exactly the same as
in \twentyfour\ through \twentyseven. Also, $T^{\r\f}_{\pm\pm}$, when
expressed in terms of $\alpha,\ldots,\delta$, is the same as in \thirty.
The relation between the free
$\chi$ and $\Omega$ fields and the ``physical" fields $\rho$ and $\phi$ are
different, however, and we must replace \twentynine\ by
\eqn\thirtysix{\eqalign{ 2\Omega(\omega) =
\gamma(x^+)+\delta(x^-)-&\alpha(x^+)-\beta(x^-)\cr
	&-{1\over\kappa}\int^{x^+}dy~e^{2\alpha(y)}\int^{x^-}dz~e^{2\beta(z)}\cr
\rho + \log(\lambda) + {\omega^2\over\kappa} -{2\over\kappa}\log\omega&  =
	\gamma(x^+)+\delta(x^-)-\half\log|\kappa|+\log(2)+\half \cr
&-{1\over\kappa}\int^{x^+}dy~e^{2\alpha(y)}\int^{x^-}dz~e^{2\beta(z)}~.}}
This formula is valid for both signs of $\kappa$. In what follows, we shall
compare the behavior of the explicit solutions, \twentynine\ and \thirtysix,
of the two theories we have studied. Just to distinguish them, we will call
them Theory I (the original case) and Theory II (the theory with decoupled
ghosts). We should perhaps warn the reader that the decoupling of the ghosts
in Theory II will turn out to be less than perfect, leaving open the question
whether some further generalization of the original theory is in order.

\newsec{Static Solutions}

The sort of question we want to ask of these (alledgedly) exact quantum
theories concerns the behavior of left-moving $f$-particle excitations sent in
from infinity (where the theory is asymptotic to the weak coupling
limit of the linear dilaton vacuum). A necessary preliminary to any quantum
discussion of scattering questions, of course, is an identification of
the vacuum state around which the scattering takes place.
An important feature of the original matter-anomaly-improved action \five\
is that the classical linear dilaton vacuum ($\rho=0$, $\phi=-\lambda \sigma$)
is still a solution. It merits the appelation of vacuum state because
the curvature is everywhere zero and there is a pseudo-translation
invariance in which the dilaton is shifted along with the spatial coordinate.
It is fairly easy to see that the linear dilaton vacuum is {\it not} a
solution of either of our exact theories. Indeed, as far as we can tell,
there is no solution which has zero curvature everywhere and no obvious
candidate for ``the" vacuum state.

Fortunately for the physical interpretation of this theory, we can find a
family of static solutions whose asymptotic behavior is governed by the
linear dilaton vacuum. To proceed, we look for restrictions on the
arbitrary functions $\alpha,\ldots,\delta$ in the general time-dependent
solutions \twentynine\ and \thirtysix\ such that the solution is static
({\it i.e.} a function of $\sigma =\half( x_+-x_-)$ only) and
asymptotic to the linear
dilaton vacuum as $\sigma\to\infty$. It is fairly easy to see that these
boundary conditions imply a) that the $\alpha,\ldots,\delta$ are all linear
functions of their arguments b) that the linear terms in $\alpha$ and $\beta$
($\gamma$ and $\delta$ resp.) are equal and opposite and
c) that the linear term in $\alpha$ is completely fixed by the linear
dilaton vacuum behavior of $\phi$. Using our freedom to rescale the
$x^\pm$-coordinate and shift its origin, we can write the most general set of
$\alpha\ldots\delta$ consistent with these conditions as
\eqn\thirtyoneprime{\eqalign{
\alpha(x^+)& =\half x^+~, \qquad
		\gamma(x^+)=\half S x^+ + \half T +\half U-
		\log2-\half +\half\log|\kappa| \cr
\beta(x^-)& = -\half x^-~, \qquad
		\delta(x^-)= -\half S x^- + \half T -\half U }}
where $S$, $T$ and $U$ are arbitrary constants. Since the solutions involve
only $\gamma+\delta$, $U$ is irrelevant and we seem to have a two-parameter
family of static solutions.

The $\rho$ and $\phi$ fields are found, in the case of Theory I, by solving
\eqn\thirtytwo{\eqalign{
2\Omega(\f) = & (S-1) \sigma  + {e^{2\sigma}\over\kappa}+ T
-2T_c  \cr
\rho(\sigma)& + {e^{-2\phi}\over\kappa}+ \log\lambda =
			S\sigma  +{e^{2\sigma}\over\kappa}+ T
}}
and, in the case of Theory II, by solving
\eqn\thirtythree{\eqalign{
2\Omega(\phi) = & (S-1) \sigma  + {e^{2\sigma}\over\kappa}
+ T-2T_c
\cr
\rho(\sigma)& + {e^{-2\phi}\over\kappa} +{2\phi\over\kappa} + \log\lambda =
                        S\sigma  +{e^{2\sigma}\over\kappa}+ T
}}
where we introduced for later convenience
\eqn\tc{T_c=-{1\over 4} \log {|\kappa|\over 4 e}
\ . }
In the limit $\sigma\to +\infty$, $\rho$ and $\phi$ approach the linear dilaton
vacuum exponentially rapidly. For Theory I, we have
\eqn\thirtyfour{
\phi\sim -\sigma + \delta_I(\sigma) e^{-2\sigma}\ , \quad
	\rho(\sigma) + \log\lambda \sim
		(\delta_I(\sigma)+{\kappa\over 8})e^{-2\sigma}~,}
where $\delta_I=-{\kappa\over 2}(S\sigma+T)$ and further corrections are of
order $e^{-4\sigma}$. For Theory II, we have
\eqn\thirtyfive{
\phi\sim -\sigma + \delta_{II}(\sigma) e^{-2\sigma}\ , \quad
        \rho(\sigma) + \log\lambda \sim
                (\delta_{II}(\sigma)+{\kappa+2\over 8})e^{-2\sigma}~,}
where $\delta_{II}=-{\kappa\over 2}((S+{2\over\kappa})\sigma+T)$
and further corrections are of order $e^{-4\sigma}$.

In a static solution, the constraints must be constant and it suffices to
evaluate them at $\sigma\sim +\infty$ where one finds
(see also \thirty ), for both theories, that
$T_{\pm\pm}^{\rho,\phi} = {\kappa\over 4}(1-2S)$.
We presume that $T_{\pm\pm}^{\rho,\phi}\ne 0$
means that there is a non-zero flux of $f$-matter at infinity (or, worse yet,
a flux of ghosts) sustaining the solution. Since we are looking for ground
states, we want solutions with zero incident flux at infinity and we must
set $S=\half$. This leaves one free parameter, $T$, in the family of static
solutions, so these ``ground states'' are not unique. This parameter is
something like the ADM mass, but we have not yet {\it constructed}
a plausible ADM mass in these new theories. (By ADM mass
we mean a conserved quantity expressible in terms of the asymptotic behavior
of the fields.)
However, it will turn out that
\eqn\mass{
M=\kappa \lambda (T-T_c)}
is the obvious candidate for the mass.

Now we turn to a description of the solutions. There are two main regimes,
$\kappa < 0$ and $\kappa > 0$, for each of the two theories and the
behavior is qualitatively different in each of the four cases. We will
discuss them in turn. In all cases, as $\sigma\to\infty$, the solution
approaches the standard linear dilaton vacuum. The only issue is what happens
as $\sigma\to -\infty$ (provided the solution does not run into a singularity
on the way).

Consider first Theory I with $N>24$.
Then by \thirtytwo, as a function of $\sigma$, $\Omega$ has a minimum at
$\sigma=\half\log {\kappa\over 4}$.
At this minimum, $2\Omega=T-T_c$. On the other
hand, as a function of $\phi$ ($\le \phi_{crit}=-\half\log\kappa$),
 one has $\Omega\ge 0$, and $\Omega$ has its minimum at $\phi=\phi_{crit}$
 where $\Omega=0$. ($\Omega$ is complex for $\phi > \phi_{crit}$.)
As a consequence, $\phi$ is well determined for all $\sigma$ if $T>T_c$,
 while  for $T<T_c$ it is not. Hence the
following picture emerges. Let first $T>T_c$. Then,
as $\sigma$ decreases from $+\infty$,
$\phi$ increases, reaches some maximum $\phi_{max}$, and then decreases
to $-\infty$ as $\sigma$ runs off to $-\infty$. The curvature blows up at
$\sigma=-\infty$, and this point is a finite proper distance away from any
finite $\sigma$: $\sigma=-\infty$ is a singular horizon. The maximum value of
$\phi$ depends on the free parameter $T$ in the general static solution. As
$T$ decreases, $\phi_{max}$ increases, and eventually at
$T=T_c$ passes through the
critical value $\phi_{crit}=-\half \log\kappa$, at which point a
(naked) curvature
singularity forms on the timelike line where $\phi=\phi_{crit}$. These
solutions are very similar to the finite mass static solutions which were
found in the study of the original action \six\ with $D(e^{2\phi})=1$.
The fact that a naked singularity is present precisely for $M<0$ confirms
 our interpretation of $M$ as the mass of the solution.

Now let $N<24$. As has been noted before, there is no critical value of $\phi$
at which a singularity is bound to occur, so these solutions should be in
some sense better-behaved. The equation for $\phi$ is best regarded as
an equation for $\omega=e^{-\phi}$ and the solution is such that, as
$\sigma$ runs from $+\infty$ to $-\infty$, $\omega$ runs monotonically from
$+\infty$ to $-\infty$. Since $\omega$ is the inverse of the coupling constant
of the theory, the $\sigma=-\infty$ side of the world has weak {\it
negative} coupling.  It also has vanishing curvature and is an infinite
proper distance away from finite $\sigma$. Since the equations of motion
only involve $\omega^2$, negative coupling is probably indistinguishable
from positive coupling, and we will provisionally regard the negative
$\omega$ half of the world as just as physical as the positive half. So,
this world has two types of weak coupling vacuum, one for positive coupling
and one for negative coupling, and the static solutions glue the two together
in a non-unique way characterised by the free parameter $T$. In all these
solutions, there is a point where $\omega\to 0$ and $\phi\to\infty$, but
nothing singular happens to any other quantity.

Now consider Theory II for $N<24$. This time no excuses are needed: as
$\sigma$ ranges from $+\infty$ to $-\infty$, $\phi$ runs monotonically
from $-\infty$ (weak coupling) to $+\infty$ (strong coupling). Curvature
is everywhere finite and vanishes asymptotically in both directions. The
$\sigma=-\infty$ side of the world is an infinite proper distance away
from any finite $\sigma$. This world has a weak-coupling and a strong-coupling
vacuum (in both of which curvature vanishes) and the static solutions glue the
two together in a non-unique way characterised by the free parameter $T$.
This class of solution seems to have a fairly straightforward physical
interepretation.

The $N>24$ solutions of Theory II seem to be rather sick:
the equation to solve for $\phi(\sigma)$ is not invertible and there is a
range of sigma where several values of $\phi$ satisfy \thirtythree. The
situation is reminiscent of a first-order phase transition and a possible
interpretation is that $\phi$ undergoes a discontinuous jump at a critical
value of $\sigma$. This jump might be smoothed out by quantum fluctuations
but, as we don't have a concrete idea of how this works, we will not consider
this case further. The static solution behavior described above has also been
obtained, in broad outline, by Strominger in his somewhat different version
of theory II \Ghost.

\newsec{Infalling Matter Solutions}

The static solutions, by construction, do not Hawking radiate. We must
now ask if they begin to radiate when they are subjected to a perturbation.
With that aim in mind, we consider the effect of an incoming
(left-moving) shock wave of massless $f$-matter on a static solution.
Such a shock wave is described by a matter stress tensor
\eqn\aone{
\tm\equiv \half \ppls f_i \ppls f_i = a\  \d (x^+-x^+_0)~.}
This is clearly not a static process but, since we have the general
time-dependent solutions, we can solve this problem exactly at the
classical level. All one has to do is to take a general solution as given
by \twentynine\ or \thirtysix\ and choose the functions
$\a,\ldots,\d$ such that the
constraint of vanishing total stress tensor, including \aone, is
satisfied. Determining the functions $\a,\ldots,\d$ only relies on the
form \thirty\ of $\tp$ and does not depend on the detailed form of the
function $\Omega$ appearing on the r.h.s of equation \twentynine\
(\thirtysix).

Naively one is tempted to impose the constraint
$T_{++}\equiv \tp  + \tm =0$. However, this cannot be a valid statement
since $T_{++}$ so defined does not transform as a tensor but rather as a
projective connection, which means that under a conformal change of coordinates
it picks up an extra term equal to $\k/2$ times the Schwarzian derivative
of the transition function. This problem is cured if we remember that the
$\left((\ppls\r)^2-\ppls^2\r\right)$ part of $\tp$
arises from the Polyakov-term $\sim \int\sqrt{g} R~ {\sq}^{-1} R$
which is non-local in a general gauge. The non-locality implies that we may
add an arbitrary function (more precisely, a projective connection)
$t_+$ to $\tp$. Alternatively, $\tp$ is obtained by
integrating the conservation equation for the stress tensor, which allows
for an arbitrary function of integration $t_+$.
This function $t_+$ should be determined by boundary conditions.
Under a conformal change of coordinates   $t_+$
will pick up $-\k/2$ times the Schwarzian derivative, so that $T_{++}+t_+$
is a true tensor\footnote{*}{
In the full quantum mechanical treatment $T_{++}$ transforms with an anomaly
equal to $\k/2-(2+N)/24=-26/24$ times the Schwarzian derivative, and one
could interpret $t_+$ as the ghost stress tensor cancelling exactly the
$-26/24$.}, and the correct constraint is
\eqn\atwo{T_{++}+t_+=0~.}
Typically we do not want to have any incoming stress-energy besides
the one specified by $\tm$, hence $t_+=0$. As we just stressed, the latter
is a coordinate-dependent statement, and it is reasonable to impose $t_+=0$
in those coordinates that are asymptotically Minkowskian (linear coordinate
transformations that do preserve the Minkowskian system have vanishing
Schwarzian derivative).
In the following we will assume that such a coordinate system has been adopted
\footnote{**}{In fact, we use rescaled asymptotically Minkowskian coordinates
such that $\r +\log \lambda\sim 0$, so that the truely asymptotically
Minkowskian coordinates are $x^\pm /\lambda$.
Note that refs. \CGHS-\ST\ call $\sigma^\pm$ the asymptotically
Minkowskian coordinates. What is called $x^\pm$ there are Kruskal-type
coordinates corresponding to our $\pm e^{\pm x^\pm}$.}.
Of course, everything said also applies to the minus components of the
stress tensor.

Combining eqs \thirty\ and \aone, we have to solve
\eqn\athree{\eqalign{
(\ppls\g)^2-(\ppls(\g-\a))^2-\ppls^2\g&={a\over \k}
\d(x^+)= {a\over \k} \ppls^2
\left( x^+ \theta(x^+)\right) \cr
(\pmin\d)^2-(\pmin(\d-\b))^2-\pmin^2\d&=0 } }
where we shifted $x^+$ in order to set $x^+_0=0$.
For $x^+\ne 0$ the r.h.s. of both equations vanish and we know that the
static solutions discussed above solve these constraints if the parameter
$S$ is set equal to one half.

To be definite,
suppose that in the far past ($\tau \to -\infty$), i.e. $x^+<0$
almost everywhere,
the $\r,\f$-system is described by one of the static solutions with
$S=\half$ but $T$ arbitrary. Now we have to
patch this particular solution to a solution for $x^+>0$, in such a way
that $\tp$ has the correct singularity at $x^+=0$, as specified by \athree,
and that $\r$ and $\f$ are continuous across the line $x^+=0$. In general,
the solution for $x^+>0$ need not be static. Also, eq. \athree\ does {\it not}
specify the solution for $x^+>0$ uniquely. Indeed, given a solution
$\a,\ldots,\d$ that satisfies \athree\ we can always change
$\a(x^+)\to\a(x^+)+\theta(x^+) g(x^+)$,
$\g(x^+)\to\g(x^+)+\theta(x^+) f(x^+)$, where e.g. $f$ and its first and
second derivatives vanish at $x^+=0$, and $g(x^+)$ is obtained by solving
$\half g'^2+(\ppls\a-\ppls\g-f')g' +\half f''-\ppls\a f'=0$ (choosing the
solution with $g'(0)=0$
and the integration constant such that $g(0)=0$).
This non-uniqueness does not, of course, contradict
the fact that the equations of motion uniquely determine all quantities in
the future once the initial data are specified. The point is that giving
$\f$ and $\r$ or $\a,\ldots,\g$ at any finite $\tau$, for $x^+<0$ only,
is not
a complete set of initial data, and there are many different solutions.
We will have to impose further physical conditions for $x^+>0$ to single
out certain solutions.

We will work out in some detail one solution that
asymptotes to the one considered in refs. \CGHS-\ST. Consider the following
choice for the functions $\a,\ldots,\d$:
\eqn\afour{\eqalign{
\a(x^+)={1\over 2}x^+\ ,\quad
&\g(x^+)={1\over 4}x^+ -{a\over \k} \left( e^{x^+}-1\right)
 \theta(x^+)+\half T-\half T_c
\cr
\b(x^-)=-{1\over 2}x^-\ ,\quad
&\d(x^-)=-{1\over 4} x^-+\half T} }
For $x^+<0$ this reproduces correctly our well-known static
solutions. It is easy to verify that the discontinuity of $\ppls\g$ is
exactly such that \athree\ is satisfied. From \twentynine\ or
\thirtysix\  we obtain
for $x^+>0$
\eqn\afive{\eqalign{
2\Omega(\f)&={1\over \k}e^{x^+}\left( e^{-x^-}-a\right)
-{1\over 4} x^+ +{1\over 4} x^- +T+{a\over \k}
-\half T_c \cr
\r(x)+\log\lambda+{1\over\k} e^{-2\f}+\left( {2\over \k}\f\right)
&={1\over\k} e^{x^+}\left( e^{-x^-}-a\right) +{1\over 4} x^+-{1\over 4}x^-
+T+{a\over \k} } }
where the term $\left( {2\over \k} \f\right)$ is present only in Theory II.
It is obvious that $\f$ and $\r$ are not static in the $x^\pm$ coordinates
(i.e. not dependent only on $\sigma=\half (x^+-x^-)$). Moreover, one can prove
that for $a\ne 0$ there is no conformal coordinate transformation that makes
$\f$ and $\r$ static for $x^+>0$. The best thing we can do is to make $\f$
and $\r$ quasi-static in a certain asymptotic region. Equations \afive\ suggest
to introduce new coordinates $\xhpm$ by
\eqn\asix{
\xhp=x^+\ ,\quad \xhm=-\log\left( e^{-x^-}-a\right)\quad {\rm for\ }
x^- <-\log a\ .}
Note that the line $x^-=-\log a$ corresponds to $\xhm=+\infty$ and thus is
a horizon (at finite proper distance as one can verify from the formula
below). These $\xhpm$ coordinates are such that for $x^+=\xhp\to\infty$, the
leading term on the r.h.s. of \afive\ is ${1\over \k} e^{\xhp-\xhm}=
{1\over \k} e^{2\hat \sigma}$ and $\f$ and $\r$ are quasi-static.
However, the next to leading terms are not static. We have for $\xhp\to
\infty$ or $\hat\sigma\to\infty$ with finite $\xhm$
\eqn\aseven{\eqalign{
\f(\hat x)&\sim -\hat\sigma +\d_I\, e^{-2\hat\sigma}\cr
\r(\hat x) +\log\lambda&\sim \left(\d_I+{\k\over 8}\right) e^{-2\hat\sigma}
}}
in Theory I and
\eqn\asevenprime{\eqalign{
\f(\hat x)&\sim -\hat\sigma +\d_{II}\, e^{-2\hat\sigma}\cr
\r(\hat x) +\log\lambda&\sim \left(\d_{II}+{\k+2\over 8}\right)
e^{-2\hat\sigma}
}}
in Theory II, where
\eqn\aeight{\eqalign{
\d_I&=-{\k\over 2}\left( \half \hat \sigma +T+{a\over \k}-{1\over 4}
\log\left( 1+ae^\xhm\right) \right)\cr
\d_{II}&=-{\k\over 2}\left( \left(\half +{2\over \k}\right)
\hat \sigma +T+{a\over \k}-{1\over 4}
\log\left( 1+ae^\xhm\right) \right)\ .}}

In general we wish
to measure Hawking radiation, if any, in asymptotically Minkow\-skian
coordinates. As we see from \aseven\ and \asevenprime,
the $\xhpm$ coordinates, or more precisely the rescaled $\xhpm/\lambda$
coordinates, are asymptotically Minkowskian.
The rate of Hawking radiation is given by the stress-energy
$t_-(\xhm/\lambda)$ flowing out to infinity. As discussed above,
we have the following transformation rule for $t_-$:
\eqn\anine{
t_-(\xhm)=\left({\partial \xhm\over \partial x^-}\right)^{-2}
\left[ t_-(x^-)+{\k\over 2} D^{\rm S}_{x^-}[\xhm]\right] }
where $D^{\rm S}$ denotes the Schwarzian derivative of $\xhm$ with
respect to $x^-$:
\eqn\aten{
D^{\rm S}_{x^-}[\xhm]={(\xhm)'''\over (\xhm)'}-{3\over 2} \left(
{(\xhm)''\over (\xhm)'}\right)^2 = - \half \left[ 1-\left( 1+ae^\xhm\right)^2
\right] \ .}
Since $t_-(x^-)=0$ and $t_-(\xhm/\lambda )=\lambda^2 t_-(\xhm)$ we
arrive at
\eqn\aeleven{
t_-(\xhm/\lambda )={1\over 4} \lambda^2 \k
\left[ 1-{1\over \left( 1+ ae^\xhm\right)^2} \right]\ .}
This is the rate of Hawking radiation in the asymptotically Minkowskian
coordinates $\xhm/\lambda$.
The reader will recognize that this is exactly the same result as found by
CGHS \CGHS, except for the replacement
${N\over 12}\to \k\equiv {N-24\over 12}$. This is not surprising since the
computation only depends, except for the prefactor $\k$, on the transformation
\asix\ which is the same as in ref \CGHS. What is different is that in ref.
\CGHS\ the transformation \asix\ was sufficient to make $\f$ and $\r$ truely
static, while in our case it is not, thanks to the
inclusion of backreaction.
As $\xhm\to \infty$, the Hawking radiation is emitted at a constant rate
${1\over 4} \lambda^2\k$ (in $\xhpm/\lambda$ coordinates),
and as $\xhm$ gets large, the total amount of
energy $E(\xhm)$ radiated away grows like
$E(\xhm)\sim {1\over 4}\lambda\k\xhm$. The same result was found in the CGHS
scenario, but there it was nonsensical since the solution was static.
At present, however, the solutions for $x^+>0$ are only quasi-static and
comparing eqs. \aeight\ with \thirtyfour\ and \thirtyfive\
we see that we can define an effective (or adiabatic) $T$-parameter $T_{\rm
eff} (\xhm)$ by
\eqn\athirteen{
T_{\rm eff} (\xhm)=T+{a\over \k}-{1\over 4} \log \left( 1+ae^\xhm\right)
\ .}
Recall that
$M=\lambda\k (T-T_c)$ is the appropriate mass
parameter of the static solutions. Thus, as $\xhm$ gets large,
\eqn\afourteen{
M_{\rm eff}(\xhm)=\lambda\k (T_{\rm eff}(\xhm)-T_c)
\sim
\lambda\k\left( T-T_c+{a\over \k}-{1\over 4}\log a\right) -{1\over 4}\lambda\k
\xhm \quad
{\rm as\ } \xhm\to\infty }
and the ``effective mass" decreases at the same rate as the Hawking
radiation carries away the energy. Thus Hawking radiation is accompanied
by the right backreaction.\footnote{*}{
We note that the approach to computing the Hawking radiation as taken e.g.
in refs \RST,\ST\
 is much the same. There, an apparent horizon (a line
where $\ppls\f=0$) is determined. Then the value of $\f$ on that horizon
is related to the mass of the solution and it is found that, for $N>0$
(corresponding in our case to $\kappa >0$) the apparent horizon recedes, hence
$\f$ increases and the mass decreases. Decreasing mass then is interpreted
 as due to outgoing Hawking radiation. In this spirit eq. \aeleven\ just
appears as a consequence of \afourteen. Note that we
derived both equations independently from each other. For $\k<0$ the situation
is just reversed. (Note that
the apparent horizon now is on the other side of the global
event horizon at $x^-=-\log a$ and approaches it as $x^+\to\infty$.) }

There are still two unsatisfactory points. First, the Hawking radiation
goes on forever (as $\xhm\to\infty$) at a rate independent of the initial state
(parametrized by $T$) and of the strength $a$ of the shock wave.
Second, and more annoying, the Hawking radiation is proportional to
$N-24$, even in Theory II, which was designed not to produce negative energy
Hawking radiation from the ghosts. For $N<24$ the Hawking radiation has
the wrong sign! Clearly, all is not well with the dynamical solutions, but
whether that is a defect of the underlying theory or the (classical)
approximation is not yet known, see however the ``Note Added" below.

\newsec{Conclusions}

What does this all mean for the usual Hawking radiation puzzles?
The initial proposal of \CGHS\ was that by adding a
matter anomaly term to the dilaton-gravity action, and then solving the
resulting {\it effective} action classically, one could expect to see the
back-reaction on the metric of the matter Hawking radiation and even hope to
address the question of the final state of black hole evolution. This
proposition seemed to be borne out by subsequent analytic and numerical
analysis of the CGHS equations with the exception that localized regions of
strong coupling, where the underlying approximation breaks down, appeared in
all ``interesting'' time evolutions.

We have attempted to improve these calculations by making
the underlying action be a $c=26$ conformal field theory, even in the strong
coupling regions, and therefore, according to the conventional wisdom, a
consistent 2D quantum gravity. To achieve this, we only had to modify the
dilaton dependence of the cosmological constant in a way that left its
weak-coupling behavior intact (which was all we had any right to believe
we understood anyway). As a bonus, we were able to construct exact classical
solutions of the modified theory. We also found that Strominger's modification
of the anomaly action, with its more sensible treatment of the ghosts, could be
handled in a similar way.

These modifications of the theory, and especially Strominger's
suggestion for decoupling the ghosts
(at least for $\k <0$), significantly improve the behavior
and physical interpretation of the static solutions of the classical
equations of motion. The dynamical solutions are another matter entirely.
Rather disappointingly, one can find solutions of the ``decoupled ghost"
theory in which the Hawking radiation rate is proportional to
$\kappa={N-24\over 12}$ (which is to say that the ghosts have not really
decoupled after all: by choosing $N<24$, one gets radiation of the wrong
sign). This problem might be solved by including graviton-dilaton quantum
loops in the calculation: all that is needed is to replace $\kappa$ in
appropriate places by $\kappa+2$, an effect that could be generated in
one-loop order (see
``Note Added" below). It might also be that some more subtle version of
Strominger's modification of the Polyakov action is called for: that
action decouples the ghosts from classical solutions of the original
classical action, but not necessarily from the solutions of the modified
actions we have been studying.

More generally, to obtain truly reliable results, a full quantum
treatment of the underlying field theory should be given. The fact that we
could
construct an {\it exact} classical solution gives hope that an exact
quantum solution can be found, but this line of inquiry has yet to be pursued.

\centerline{\bf Acknowledgements}

This work was supported in part by DOE grant DE-FG02-91ER40671. We would
like to thank D. Kutasov for an illuminating discussion on exactly soluble
dilaton-gravity models. We would also like to thank L.~Thorlacius for
numerous helpful suggestions.

\vskip 5.mm
\centerline{\bf Note Added}

While preparing this paper, we received
a preprint by S. de Alwis in which similar observations to ours are made.
After submitting this paper to hepth@xxx, we learned that S. Giddings and
A. Strominger also made similar progress  towards an exact $c=26$ conformal
theory, and that the same idea was also contained in ref. \TSEY\ by J. Russo
and A. Tseytlin without however giving the explicit free field realization
or writing the general solution in closed form. Also,
two-dimensional dilaton gravities, although with a somewhat
different action, have been studied in the past by
A. Chamseddine \CHAM, and more recently by
T.Burwick and A. Chamseddine \CHAM.

For completeness, we would like to include here two observations made
after submitting this paper for publication:

First, we would like to argue that a full quantum treatment most
probably replaces the $\kappa$ prefactor in the Hawking radiation rate
by $N/12$, thus making it manifestly positive for all $N$.
The basic ingredients in the computation of the Hawking radiation rate were the
$T_{\pm \pm}=0$ constraints
$$T_{\pm\pm}\equiv
T_{\pm \pm}^{\rho,\phi}+T_{\pm\pm}^{\rm M}+T_{\pm\pm}^{\rm gh}
+t_\pm =0
$$
(where we now explicitly include the ghost contribution)
and the anomalous transformation law of $T^{\rho,\phi}$. Indeed,
in the classical theory (based on our quantum-improved action) only
$T_{\pm\pm}^{\rho,\phi}$ transforms anomalously. Its transformation law is
$$T_{\pm\pm}^{\rho,\phi}(y^\pm)=
\left( {\partial y^\pm\over \partial x^\pm}\right) ^{-2}
\left(
T_{\pm\pm}^{\rho,\phi}(x^\pm) + { c^{\rho,\phi}_{\rm cl}\over 24}
D^S_{x^\pm} [y^\pm] \right)$$
where the anomaly is proportional to the classical central charge
(i.e. the one obtained by a Poisson bracket computation)  of the
$\phi, \rho$ -system: $c_{\rm cl}^{\rho,\phi}=-12\kappa\ .$
This makes it necessary to introduce $t_\pm$ to cancel this anomaly
(cf (7.10)) in order that $T_{\pm\pm}=0$ be a coordinate-invariant statement.

In the quantum theory however, $c^{\rho,\phi}=2-12\kappa$,
and $T_{\pm\pm}^{\rm M}$ and $T_{\pm\pm}^{\rm gh}$ transform
{\it with} anomalies given by $c^{\rm M}=N$ and $c^{\rm gh}=-26$
so that the total stress tensor
$T_{\pm\pm}-t_\pm$ transforms {\it without} anomaly :$c^{\rm
tot}=c^{\rho,\phi}+c^{\rm M}+c^{\rm gh}=0$. There is now no need for
a $t_\pm$ term and we do not include one: the full energy-momentum tensor
is now defined to be
$T_{\pm \pm}^{\rho,\phi}+T_{\pm\pm}^{\rm M}+T_{\pm\pm}^{\rm gh}$.
We also make the physically very reasonable assumption that all
energy flowing in or out from infinity is described by the matter
energy-momentum tensor $T_{\pm\pm}^{\rm M}$ (the ghosts are unphysical
and the graviton/dilaton degrees of freedom don't propagate:
only matter remains). In particular, outgoing Hawking radiation should
be signalled by a non-vanishing {\it matter}
stress energy tensor $T_{--}^{\rm M}$.

Now, the matter energy-momentum tensor, taken by itself, {\it does} have an
anomalous transformation law. According to the above discussion,
in a transformation between coordinates $x$ and $\hat x$ we have
$$T_{--}^{\rm M}(\hat x^-)=
\left( {\partial \hat x^-  \over \partial x^-  }\right) ^{-2}
\left(
T_{--    }^{\rm M }(x^-  ) + {N\over 24}
D^S_{x^-  } [\hat x^-] \right)\ . $$
We can apply this to the infall solution described in Sect. 7 and in
particular to the coordinate transformation of \asix . The value
of $T_{--}^{\rm M}$ in the $x$ coordinates is clearly zero, but we
would like to measure outgoing matter energy in the asymptotically
Minskowskian coordinates $\hat x$. The Schwarzian derivative $D^S$ is
still given by \aten\ and we find
$$
T_{--}^{\rm M}(\hat x^-)= {N\over 48}\left[
1-{1\over \left( 1+ae^{\hat x^-}\right)^2}\right]\ .$$
(Recall that the truely asymptoyically Minkowskian coordinates are
$\hat x^\pm /\lambda$ and that $T_{--}(\hat x^-/\lambda)=
\lambda^2 T_{--}(\hat x^-)$.)
Thus we find that the Hawking radiation rate is proportional to $N$
as it should and that it is given by exactly the same expression
as in CGHS [1].

Second, we would like to comment on the issue of eternally continuing
Hawking radiation and negative mass solutions.
At the level of solutions of the equations of motion of our action
(4.1) we found that, for $\k>0$, as the Hawking radiation proceeds the
effective mass $M_{\rm eff}$ eventually becomes negative. This precisely
occurs when $T_{\rm eff}=T_c$.
Now, we have seen for the static solutions that
at this point a naked time-like singularity occurs. This means that the
subsequent evolution is not well defined unless new boundary conditions
are imposed on the line of singularity.
This situation has been discussed in detail by
Russo, Susskind and Thorlacius \RSTE\ who consider the same action (4.1) in
terms of $\Omega$ and $\chi$ but with a slightly modified relation between
$\Omega,\chi$ and $\phi,\rho$. Their analysis carries over to our case,
word by word.
 One finds that at the point where the apparent horizon
(defined by $\partial_+\Omega = 0$)
intersects the line of singularity
(defined by $\Omega=0$),
the $\Omega,\chi$-solution relevant
to the infalling shock wave scenario can be matched continuously onto the
static $M=0$ (i.e. $T=T_c$) solution.
More precisely, before the emergence of the naked singularity, the effective
mass is given by (cf \athirteen )
$$M_{\rm eff}= M + a \lambda-{1\over 4}\kappa\lambda\log\left(1+
a e^{\xhm}\right)
$$
with $\xhm$ given by \asix, and where $M=\kappa\lambda (T-T_c)$
is the mass of the the
``initial" (static) solution. At $\xhm=\xhm_i$ where the line of singularity
intersects the apparent horizon, $M_{\rm eff}$ is zero, and by the above
choice, $M_{\rm eff}$ remains zero for $\xhm >\xhm_i$. Since we have
incorporated backreaction consistently, the Hawking radiation must stop when
$M_{\rm eff}=0$ is reached. (Of course, at $\xhm =\xhm_i$ there is also the
thunderpop \RSTE.) We see that the occurrence of negative mass solutions from
a positive (or zero) mass initial state is eliminated by appropriate
boundary conditions. This is very satisfactory at the level of the
equations of motion, although, at present, we do not know how to deal with this
at the quantum level.

\listrefs
\bye